\documentclass[runningheads]{llncs}
\usepackage[T1]{fontenc}
\usepackage{cite}
\usepackage{amsmath,amssymb,amsfonts}
\usepackage{algorithmic}
\usepackage{graphicx}
\usepackage{textcomp}
\usepackage[dvipsnames,table]{xcolor} 
\usepackage{soul} 
\usepackage{lipsum}
\usepackage{multirow}
\usepackage{hyperref}
\usepackage{colortbl}
\usepackage{subfig}
\usepackage{orcidlink}
\usepackage{caption}
\usepackage{lettrine}  
\usepackage{setspace}
\usepackage{booktabs} 
\usepackage{siunitx}  
\usepackage{array}     

\begin{document}

\title{Low Power Approximate Multiplier Architecture for Deep Neural Networks}

\author{
Pragun Jaswal\textsuperscript{1}~\orcidlink{0009-0005-9580-3184} \and
L. Hemanth Krishna\textsuperscript{2}~\orcidlink{0000-0002-7737-6685} \and
B. Srinivasu\textsuperscript{3}~\orcidlink{0000-0003-0974-8245}
}

\authorrunning{Pragun Jaswal et al.}

\institute{
School of Computing and Electrical Engineering,\\
Indian Institute of Technology Mandi, Mandi - 175005, India\\
\textsuperscript{1}\email{thepragun@gmail.com}, 
\textsuperscript{2}\email{hemanthkrishna412@gmail.com}, 
\textsuperscript{3}\email{srinivasu@iitmandi.ac.in}
}

\maketitle
\begin{abstract}
This paper proposes an low power approximate multiplier architecture for deep neural network (DNN) applications. A 4:2 compressor, introducing only a single combination error, is designed and integrated into an 8×8 unsigned multiplier. This integration significantly reduces the usage of exact compressors while preserving low error rates. The proposed multiplier is employed within a custom convolution layer and evaluated on neural network tasks, including image recognition and denoising. Hardware evaluation demonstrates that the proposed design achieves up to 30.24\% energy savings compared to the best among existing multipliers. In image denoising, the custom approximate convolution layer achieves improved Peak Signal-to-Noise Ratio (PSNR) and Structural Similarity Index Measure (SSIM) compared to other approximate designs. Additionally, when applied to handwritten digit recognition, the model maintains high classification accuracy. These results demonstrate that the proposed architecture offers a favorable balance between energy efficiency and computational precision, making it suitable for low-power AI hardware implementations.

\end{abstract}
\section{Introduction}
\vspace{1em}
\lettrine[lines=2, lhang=0, loversize=0.1]{\textbf{C}}{ onvolutional} Neural Networks (CNNs) have become fundamental in advancing fields such as computer vision, speech recognition, and multimedia processing, driven by their ability to extract complex features from large datasets. With the rapid expansion of real-time and edge-based applications, particularly in resource-constrained environments, the demand for highly efficient hardware implementations has intensified. Traditional exact arithmetic circuits impose significant overheads in terms of area, delay, and power consumption\cite{fahim2002}, making them less favorable for large-scale data processing tasks and edge-based applications~\cite{Jiang}. To deal with these issues and for applications which involve human perception, approximate computing has evolved as a viable alternative \cite{Chippa}, allowing slight accuracy trade-offs to accumulate large advantages in energy, area and speed ~\cite{Bosio2022,Liu}.

Approximation techniques, particularly in arithmetic units such as adders, multipliers, have shown considerable success \cite{Soares,Gupta,Shafique}. Among these, the use of approximate compressors in the partial product reduction (PPR) stage of Dadda's multipliers has gained significant attention. Approximate compressors simplify the logic by approximating low-probability signal combinations, reducing both the size and error probability of multipliers. 

The fundamental motivation of this work is to introduce approximations selectively in combinations with the lowest occurrence probability $P(1/256)$, thereby significantly reducing both the hardware complexity and the error probability of the multiplier. In this study, we specifically focus on the design and analysis of approximate 4:2 compressors and 8 bit unsigned multiplier. A comprehensive survey of existing 4:2 compressor architectures was conducted. Based on this analysis, we propose a high-accuracy 4:2 approximate compressor design that introduces only a single error combination while achieving a 30.24\% improvement in energy consumption compared to the best existing designs.\\


\noindent\textbf{The main contributions of this paper are as follows:}
\begin{enumerate}
    \item A 4:2 approximate compressor is proposed, integrated into a high-accuracy 8×8 multiplier, achieving up to 8.93\% energy savings and improving power efficiency.
    \item The proposed multiplier achieves up to 27.64\% power and 27.48\% energy reduction over the best existing multiplier-1 design.
    \item The proposed multiplier achieves up to 33.14\% power and 30.24\% energy reduction over the best existing multiplier-2 design.
    \item The multiplier is integrated into a custom convolution layer for DNN tasks, such as image denoising and digit recognition, demonstrating high accuracy with reduced computational overhead.
\end{enumerate}


\section{Related Work}
The exact 4:2 compressor takes four primary inputs $x_1$, $x_2$, $x_3$, and $x_4$ along with an additional carry input $C_{in}$. These five inputs are summed to produce an output that can range from 0 to 5.
To represent this maximum sum of 5, the compressor requires three output bits. These outputs are $Cout$, $Carry$ which has a weight of $2^{n+1}$ and $Sum$ with a weight of same as inputs $2^n$.
Each output contributes to reconstructing the total sum based on its positional weight in binary representation. This structure enables efficient reduction of partial products in arithmetic circuits, especially in multiplier architectures. 

\begin{figure}[ht]
    \centering
    \includegraphics[width=0.30\linewidth]{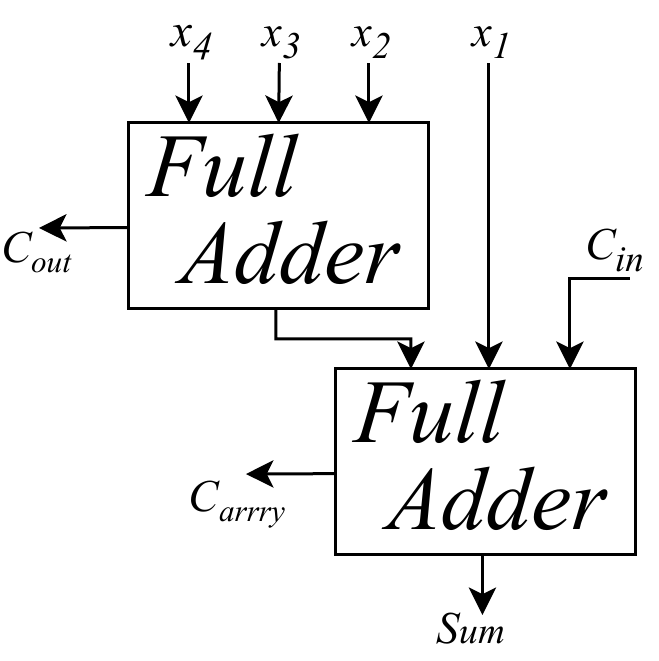}
    \caption{Conventional Exact 4:2 Compressor \cite{amir2015} }
    \label{fig:Exact}
\end{figure}

Fig.~\ref{fig:Exact} illustrates the implementation of an exact 4:2 compressor using full adders. In contrast, the approximate 4:2 compressor typically eliminates the $Cin$ and $Cout$ pins, computing only ($Carry$, $Sum$), as the sum is $x_1 + x_2 + x_3 + x_4$. This elimination breaks the carry propagation chain between compressors, thereby accelerating the accumulation of the sum.
However, the maximum value that can be encoded using only the Sum and Carry outputs is three. Given four input bits ($x_1$ to $x_4$), it is evident that at least one error is unavoidable, specifically when all inputs are ‘1’.

\subsection{Low Accuracy Approximate 4:2 Compressors}
Most researchers have proposed various approximate 4:2 compressor designs with different combinations of error cases. For instance, if four input combinations produce incorrect outputs, the error rate is 25\%~\cite{amir2015,hemanth2021,Hwang2025}. When five combinations result in errors, the error rate increases to 31.25\%~\cite{hemanth2024}, and with six erroneous combinations, it rises to 37.5\%~\cite{Zhang2023,Yongqiang2023}.

These designs primarily aim to reduce the hardware complexity 
of compressor circuits. 
However, this reduction often comes at the cost of accuracy. 
When integrated into an 8-bit multiplier architecture,
such approximate compressors tend to 
exhibit higher error metrics,
particularly in terms of Normalized Mean Error Distance (NMED) 
and Mean Relative Error Distance (MRED), which are further discussed in Section IV.

The 4:2 compressor design proposed in \cite{Anil2023} uses 
two XOR gates for the $Sum$ output, introducing up to four combination errors,
which results in an error probability of $ P(16/256)$ , 
as shown in Table \ref{tab:hardware_synthesis}.
The compressors proposed in \cite{hemanth2024} 
use an input reordering circuit with additional gates, 
eliminating XOR gates in the critical path, and introduce two combination errors,
resulting in an error probability of \( P(19/256) \). 
In \cite{Anupam2025}, two-compressor are proposed, where Design-2, uses only OR and AND gates, incurs a maximum of seven error combinations with an error probability of  \( P(55/256) \) . The compressor design in
\cite{Zhang2023} incorporates one XOR and one NOR gate in the critical path, 
introducing a maximum of six combination errors, leading to an error probability
of \( P(70/256) \). These low-accuracy compressors provide substantial energy savings, though at the expense of reduced computational accuracy.

\subsection{High-Accuracy Approximate 4:2 Compressors}

Compressors with a single combination error typically occur 
when all inputs are logic high. It corresponds to an error probability of 1/256 
and provides significantly improved accuracy in multiplier architectures.
In this work, such designs are classified as \textit{high-accuracy approximate compressors}.
Examples of such designs are found in \cite{Anupam2025, Strollo2020, Yang2015, Kong2021},
where the proposed compressors maintain low error rates. However,
these designs often require higher hardware resources and introduce 
longer critical paths, which can lead to increased delay and power consumption.


\section{Proposed Approximate Multiplier Architecture}
This section is divided into two parts. The first describes the construction of the 8×8 approximate unsigned multiplier architecture incorporating the proposed compressor. The second presents the design of the proposed 4:2 approximate compressor, which serves as the core component of the multiplier.

\subsection{Efficient Architecture of Approximate Multiplier}
\begin{figure*}

    \centering
    \includegraphics[width=1\linewidth]{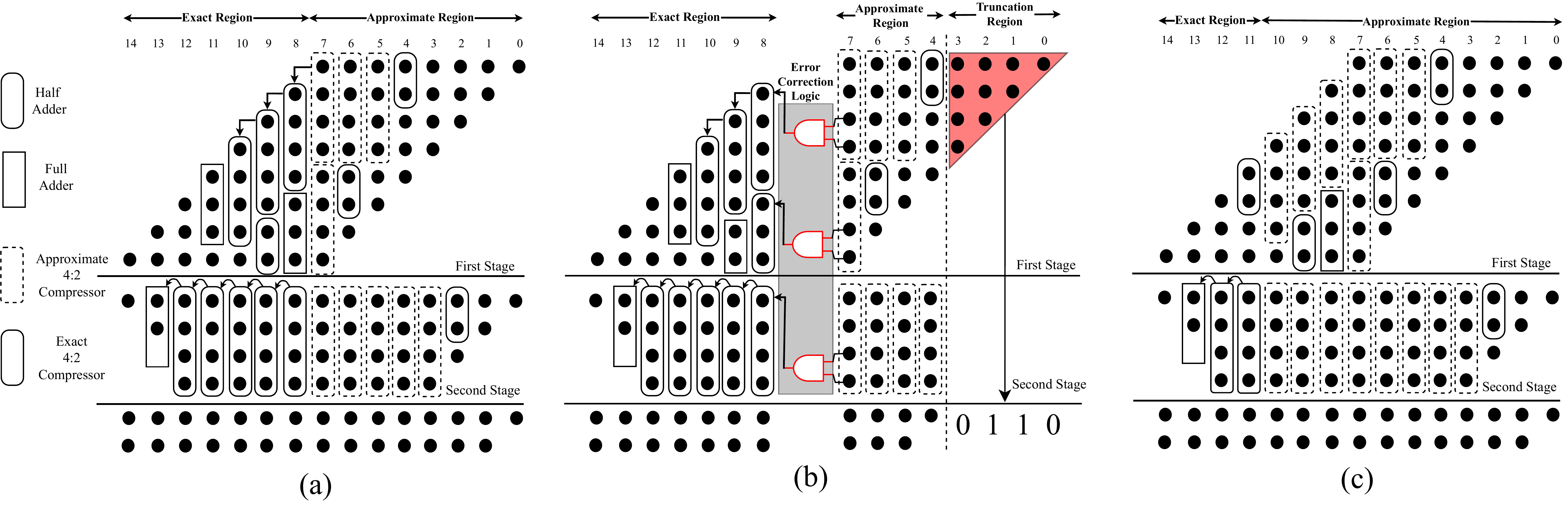}
    \caption{Structure of $8 \times 8$ Unsigned Multipliers (a) Multiplier Design-1 : from \cite{hemanth2024}, \cite{Strollo2020}, \cite{Kong2021}, (b) Multiplier Design-2 : from \cite{Zhang2023}, \cite{Anil2023}, (c) Proposed multiplier for High Accuracy Compressor}
    \label{fig:Dot Diagram}
\end{figure*}

Previous multiplier designs, shown in Fig.~\ref{fig:Dot Diagram}~(a,~b), use a mix of exact and approximate compressors. Typically, exact compressors are used in the most significant columns to maintain accuracy, while approximate compressors are applied in the least significant columns to reduce hardware cost. Additionally, design (b) incorporates truncation in the least significant columns $n-4$ and an error correction module to further mitigate the error rate.

In contrast, the proposed multiplier design as shown in Fig.~\ref{fig:Dot Diagram}~(c) uses only approximate compressors throughout the design to further reduce hardware complexity. Although it relies more heavily on approximation, the overall error remains low. This is because the proposed compressor introduces only one combination error with a very low probability of occurrence $P(1/255)$. As a result, error metrics such as NMED and MRED show only minimal increase, ensuring that the accuracy of the multiplier is not significantly affected.

\subsection{Proposed High-Accuracy 4:2 Compressor}
\vspace{-7mm}

\begin{table}[h]
    \centering
    \small
    \caption{Truth Table of the Proposed 4:2 Approximate Compressor}
    \vspace{2mm}
    \label{tab:truth table}
    \renewcommand{\arraystretch}{1.2} 
    { 
    \begin{tabular}{|cccc|c|c|cc|c|c|}
        \hline
        \multicolumn{4}{|c|}{\textbf{Inputs}} & \textbf{Exact} & \multirow{2}{*}{\textbf{Prob.}} & \multicolumn{2}{c|}{\textbf{Outputs}} & \textbf{Appr.} & \multirow{2}{*}{\textbf{Difference}} \\
        \cline{1-4} \cline{7-8}
        $x_4$ & $x_3$ & $x_2$ & $x_1$ & \textbf{Value} &  & \textbf{Carry} & \textbf{Sum} & \textbf{Value} &  \\
        \hline
        0 & 0 & 0 & 0 & 0 & $81/256$ & 0 & 0 & 0 & 0 \\
        \hline
        0 & 0 & 0 & 1 & 1 & $27/256$ & 0 & 1 & 1 & 0 \\
        \hline
        0 & 0 & 1 & 0 & 1 & $27/256$ & 0 & 1 & 1 & 0 \\
        \hline
        0 & 0 & 1 & 1 & 2 & $9/256$  & 1 & 0 & 2 & 0 \\
        \hline
        0 & 1 & 0 & 0 & 1 & $27/256$ & 0 & 1 & 1 & 0 \\
        \hline
        0 & 1 & 0 & 1 & 2 & $9/256$  & 1 & 0 & 2 & 0 \\
        \hline
        0 & 1 & 1 & 0 & 2 & $9/256$  & 1 & 0 & 2 & 0 \\
        \hline
        0 & 1 & 1 & 1 & 3 & $3/256$  & 1 & 1 & 3 & 0 \\
        \hline
        1 & 0 & 0 & 0 & 1 & $27/256$ & 0 & 1 & 1 & 0 \\
        \hline
        1 & 0 & 0 & 1 & 2 & $9/256$  & 1 & 0 & 2 & 0 \\
        \hline
        1 & 0 & 1 & 0 & 2 & $9/256$  & 1 & 0 & 2 & 0 \\
        \hline
        1 & 0 & 1 & 1 & 3 & $3/256$  & 1 & 1 & 3 & 0 \\
        \hline
        1 & 1 & 0 & 0 & 2 & $9/256$  & 1 & 0 & 2 & 0 \\
        \hline
        1 & 1 & 0 & 1 & 3 & $3/256$  & 1 & 1 & 3 & 0 \\
        \hline
        1 & 1 & 1 & 0 & 3 & $3/256$  & 1 & 1 & 3 & 0 \\
        \hline
        \rowcolor[gray]{0.8} 
        1 & 1 & 1 & 1 & 4 & $1/256$  & 1 & 1 & 3 & -1 \\
        \hline
    \end{tabular}
    }
\end{table}

\begin{figure}[h]
    \centering
    \includegraphics[width=0.6\linewidth]{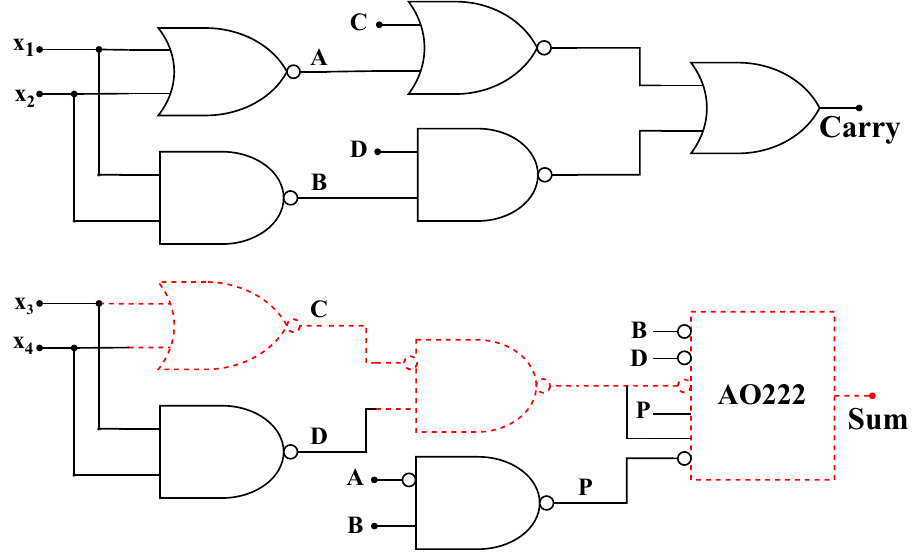}
    \caption{Proposed Approximate 4:2 Compressor with One Error Probability}
    \label{fig:Proposed Compressor}
\end{figure}

The proposed approximate compressor comprises four inputs and two outputs, defined by Equation (1) and (2) for the Carry and Sum, respectively. Equation (3) represents the intermediate values where \(A\) and \(C\) represent the NOR operations for specific input pairs and outputs \(B\) and \(D\) are likewise NAND operations. Given that NOR and NAND gates exhibit superior speed and energy efficiency compared to the AND and OR gates utilized in earlier proposed designs, this approach is well-suited for low-power, high-speed applications.

\begin{equation}
\text{Carry} = \overline{B \cdot D} + \overline{A + C} \tag{1}
\end{equation}
\begin{equation}
\begin{split}
\text{Sum} = &\; \overline{A} \cdot B \cdot C + \overline{A} \cdot B \cdot \overline{D} + \overline{A} \cdot \overline{C} \cdot D \\
             &+ \overline{B} \cdot \overline{C} \cdot D + \overline{B} \cdot \overline{D}
\end{split}
\tag{2}
\end{equation}
\begin{equation*}
\tag{3}
\text{where } \quad A = \overline{x_1 + x_2}, \quad B = \overline{x_1 \cdot x_2}
\end{equation*}
\begin{equation*}
\hspace{40px} C = \overline{x_3 + x_4}, \quad D = \overline{x_3 \cdot x_4}.
\end{equation*}

Table \ref{tab:truth table} represents the Sum and Carry outputs of the proposed approximate compressor for all possible combinations. 
The proposed compressor not only have high accuracy but also has shorter critical path.  As shown in Fig.\ref{fig:Proposed Compressor}, the red colored dotted lines marks the critical path of the approximate compressor. There are one NOR-2, one NAND-2, two inverters, and one AO222 on the critical path. The proposed compressor architecture demonstrates a notable reduction in propagation delay compared to the designs presented in \cite{Strollo2020} and \cite{Kong2021}. These performance improvements are discussed in detail in the subsequent results section.

\section{Results and Discussion}

\subsection{Error Metrics}

To evaluate the accuracy of the proposed approximate design, error metrics such as \textit{Error Distance }, \textit{Relative Error Distance }, and \textit{Mean Relative Error Distance } are used. These metrics quantify the deviation of approximate outputs from their exact counterparts.

\subsection*{Error Distance (ED)}
The Error Distance measures the absolute difference between the exact and approximate outputs for each test case as shown in Equation~(\ref{eq:ed}).
\begin{equation}
\text{ED}_i = |A_i - A'_i|
\label{eq:ed}\tag{4}
\end{equation}

where \( A_i \) is the exact output and \( A'_i \) is the approximate output for the \(i^{\text{th}}\) test case.

\subsection*{ Error Rate (ER)}
The Error Rate indicates the percentage of test cases where the approximate output differs from the exact output. It is computed as shown in Equation~(\ref{eq:er}).
\begin{equation}
\text{ER} = \left( \frac{1}{N} \sum_{i=1}^{N} \delta_i \right) \times 100
\label{eq:er}\tag{5}
\end{equation}

\[
\delta_i =
\left\{
\begin{array}{ll}
1, & \text{if } A_i \neq A'_i \\
0, & \text{if } A_i = A'_i
\end{array}
\right\}
\]

Here, \( A_i \) and \( A'_i \) denote the exact and approximate outputs respectively for the \( i^{\text{th}} \) test case, and \( N \) is the total number of test cases.

\subsection*{Relative Error Distance (RED)}
The Relative Error Distance normalizes the error with respect to the exact output as shown in Equation~(\ref{eq:red}).

\begin{equation}
\text{RED}_i = \frac{|A_i - A'_i|}{|A_i|}
\label{eq:red}\tag{6}
\end{equation}

\subsection*{Mean Relative Error Distance (MRED)}
The Mean Relative Error Distance provides the average RED across all test cases, giving an overall measure of the accuracy degradation due to approximation as shown in Equation~(\ref{eq:mred}).

\begin{equation}
\text{MRED} = \frac{1}{N} \sum_{i=1}^{N} \frac{|A_i - A'_i|}{|A_i|}
\label{eq:mred}\tag{7}
\end{equation}

where \( N \) is the total number of test cases.

\begin{table}[h]
    \centering
    \small
    \caption{Error Metrics of Proposed Multiplier Design}
    \vspace{2mm}
    \label{tab:Proposed_Multiplier_ErrorMetrics}
    \renewcommand{\arraystretch}{1.2}
    \setlength{\tabcolsep}{12pt}
    \begin{tabular}{lccc}
        \toprule
        \toprule
        \textbf{Design} & \textbf{ER (\%)} & \textbf{NMED (\%)} & \textbf{MRED (\%)} \\
        \midrule
        Design \cite{hemanth2024}      & 68.498  & 0.596 & 3.496  \\
        Design \cite{Anil2023}           & 65.425  & 0.673 & 3.531  \\
        Design \cite{Anupam2025}       & 6.994   & 0.046 & 0.109  \\
        Design \cite{Anupam2025}       & 86.326  & 1.879 & 9.551  \\
        Design \cite{Strollo2020}      & 21.296  & 0.162 & 0.578  \\
        Design \cite{Strollo2020}      & 6.994   & 0.046 & 0.109  \\
        Design \cite{Kong2021}         & 6.994   & 0.046 & 0.109  \\
        Design \cite{Kong2021}         & 6.994   & 0.046 & 0.109  \\
        Design \cite{Zhang2023}          & 95.681  & 1.565 & 20.276 \\
        Design \cite{Yang2015}         & 6.994   & 0.046 & 0.109  \\
        Proposed                         & 6.994   & 0.046 & 0.109  \\
        \bottomrule
        \bottomrule
    \end{tabular}
\end{table}

All designs of $8\times8$ unsigned approximate multipliers were evaluated by simulation across the complete input space. The corresponding error metrics for each design are summarized in Table~\ref{tab:Proposed_Multiplier_ErrorMetrics}. Although the proposed multiplier employs fully approximate compressors, it achieves a high level of computational accuracy, with a MRED of 0.109\%.

\subsection{Hardware Synthesis Analysis}

The proposed and existing 8-bit unsigned multipliers were implemented using Verilog HDL and synthesized using Cadence Genus with UMC 90nm technology, under typical-typical (TT) process conditions, to evaluate their design efficiency and characteristics. Table~\ref{tab:hardware_synthesis} presents a comparative analysis of both low and high accuracy compressor designs. The proposed compressor demonstrates an energy reduction of 9.81\% compared to the best performing high-accuracy compressor \cite{Anupam2025}.

\sethlcolor{red!30} 
\newcommand{\hlg}[1]{\sethlcolor{green!40}\hl{#1}\sethlcolor{red!50}} 

\begin{table}
    \centering
    \small
    \caption{Hardware Synthesis Metrics of Existing and Proposed 4:2 Compressors}
    \vspace{2mm}
    \renewcommand{\arraystretch}{1.2}
    \setlength{\tabcolsep}{5pt}
    \label{tab:hardware_synthesis}
    \begin{tabular}{clccccc}
        \toprule
        \multirow{2}{*}{\textit{\textbf{S.No}}} & \multirow{2}{*}{\textit{\textbf{Design}}} & \textbf{Area}  & \textbf{Power}  & \textbf{Delay}  & \textbf{PDP}  & \textbf{Error} \\
        &  &($\mu m^2$)&($\mu$W)&(ps)&(fJ)& \textbf{Probability} \\
        \midrule
        1. & Exact                       & 43.90       & 1.99 & 436 & 0.867 & 0 \\
        2. & Design-1 \cite{Yang2015}    & 50.17  & 2.39 & 469 & \hl{0.852} & 1/256 \\
        3. & Design-1 \cite{Kong2021}    & 44.68       & 1.86 & \hl{383} & 0.713 & 1/256 \\
        4. & Design-5 \cite{Kong2021}     & \hlg{28.22}      & 1.17 & 297 & 0.347 & 1/256 \\
        5. & Design-1 \cite{Anupam2025}   & 34.49      & 1.20 & \hlg{226} & 0.291 & 1/256 \\
        6. & Design-3\cite{Strollo2020}   & \hl{76.82} & \hl{3.02} & 307 & 0.827 & 1/256\\
        7. & Design-1 \cite{hemanth2024}  & 49.74      & 1.83 & 374 & 0.684 & 19/256 \\
        8. & Design \cite{Anil2023}       & 25.87      & 1.02 & 175 & 0.179 & 16/256 \\
        9. & Design-2 \cite{Anupam2025}   & 19.60      & 0.71 & 104 & 0.074 & 55/256 \\
        10. & Design-2 \cite{Strollo2020} & 31.36      & 1.37 & 308 & 0.422 & 4/256 \\
        11.& Design \cite{Zhang2023}      & 14.11      & 0.52 & 139 & 0.072 & 70/256 \\
        12. & Proposed                    & 30.57      & \hlg{1.12} & 237 & \hlg{0.265} & 1/256 \\
        \bottomrule
    \end{tabular}

    \vspace{0.3mm}
{\footnotesize *Best and worst results for the high-accuracy compressor are highlighted in green and red, respectively.}
\end{table}


\begin{table}[h]
    \centering
    \caption{Hardware Synthesis and Error Metrics (MRED, Power, Delay, PDP) of 8-bit Existing and Proposed Approximate Multipliers}
    \label{tab:Multipliers_APDP}
    \renewcommand{\arraystretch}{1.5}
    \setlength{\tabcolsep}{5pt}
    \resizebox{\textwidth}{!}{
    \begin{tabular}{@{}c|| cccc| cccc| cccc@{}}
        \toprule
        \multirow{3}{*}{\textbf{Design}} & 
        \multicolumn{4}{c|}{\textbf{Multiplier Design-1} \cite{hemanth2024,Strollo2020,Kong2021}} & 
        \multicolumn{4}{c|}{\textbf{Multiplier Design-2} \cite{Anil2023,Zhang2023}} & 
        \multicolumn{4}{c}{\textbf{Proposed Multiplier Design}} \\
        \cmidrule(lr){2-5} \cmidrule(lr){6-9} \cmidrule(l){10-13}
        & \textbf{MRED } & \textbf{Power } & \textbf{Delay } & \textbf{PDP } &
          \textbf{MRED } & \textbf{Power } & \textbf{Delay } & \textbf{PDP } &
          \textbf{MRED } & \textbf{Power } & \textbf{Delay } & \textbf{PDP } \\

         &(\%)&($\mu W$)&(ns)&(fJ)&(\%)&($\mu W$)&(ns)&(fJ)&(\%)&($\mu W$)&(ns)&(fJ)\\ 
        \midrule
        \midrule
        Design \cite{hemanth2024} & 0.993 & 76.25 & 2.084 & 158.91 & 1.286 & 74.68 & 2.009 & 150.05 & 3.496 & 63.17 & 2.042 & \textbf{129.09} \\
        Design \cite{Anil2023}      & 0.773 & 68.67 & 1.998 & 137.24 & 0.974 & 67.58 & 1.996 & 134.87 & 3.531 & 57.41  & 2.042 & \textbf{117.23} \\
        Design \cite{Anupam2025}  & 0.023 & 68.67 & 2.071 & 142.23 & 0.715 & 66.81 & 2.071 & 138.35 & 0.109 & 57.50  & 2.121 & \textbf{121.96} \\
        Design \cite{Anupam2025}  & 2.693 & 59.00 & 1.993 & 117.75 & 2.704 & 59.38 & 1.993 & 118.32 & \hl{9.551} & 41.12  & 2.042 & \textbf{\hlg{83.97}} \\
        Design \cite{Strollo2020} & 0.090 & 74.94 & 2.084 & 156.15 & 0.702 & 77.44 & 2.085 & 161.40 & 0.578 & 69.21  & 2.126 & \textbf{147.14} \\
        Design \cite{Strollo2020} & 0.023 & 97.30 & 2.239 & 217.86 & 0.715 & 78.83 & 2.140 & 168.70 & \hlg{0.109} & 82.65  & 2.189 & \textbf{\hl{180.92}} \\
        Design \cite{Kong2021}    & 0.023 & 76.94 & 2.243 & 172.54 & 0.715 & 75.30 & 2.243 & 168.82 & 0.109 & 74.13  & 2.293 & \textbf{169.98} \\
        Design \cite{Kong2021}    & 0.023 & 61.73 & 2.090 & 135.18 & 0.715 & 62.42 & 2.090 & 130.46 & 0.109 & 66.10  & 2.139 & \textbf{141.39} \\
        Design \cite{Zhang2023}     & 4.399 & 61.73 & 1.993 & 123.06 & 4.320 & 65.51 & 1.995 & 130.73 & \hl{20.276} & 42.46 & 2.042 & \textbf{\hlg{86.70}} \\
        Design \cite{Yang2015}    & 0.023 & 70.19 & 2.350 & 164.94 & 0.715 & 71.25 & 2.350 & 167.44 & 0.109 & 62.69  & 2.371 & \textbf{148.64} \\
        \textbf{Proposed}           & 0.023 & 65.56 & 1.993 & 130.75 & 0.715 & 64.25 & 1.993 & 128.06 & \hlg{0.109} & 44.66  & 2.042 & \textbf{\hlg{91.20}} \\
        \bottomrule
    \end{tabular}
    }

    \vspace{0.3mm}
{\footnotesize *Best and worst results for the are highlighted in green and red, respectively.}
\end{table}

\begin{figure}[h]
    \centering
    \includegraphics[width=0.7\linewidth]{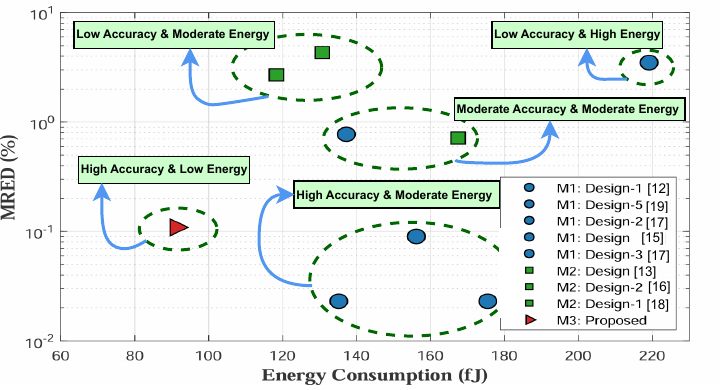}
    \caption{Comparison of PDP and MRED for Different Designs}
    \label{fig:PDP and MRED comparison}
\end{figure}

Furthermore, the proposed multiplier achieves the lowest Power-Delay Product (PDP) among all designs, with a value of 91.20 fJ as presented in Table \ref{tab:Multipliers_APDP}. Proposed design demonstrates significant energy improvement as depicted in Fig.~\ref{fig:PDP and MRED comparison}, with reductions of 27.48\% and 30.24\% in energy consumption compared with best of multiplier design-1 and design-2, respectively. When high-accuracy compressors, described in \cite{Strollo2020} and \cite{Kong2021}, are employed within the proposed multiplier, the proposed compressor achieves notable energy improvements of 46.35\% and 49.59\%, respectively.

\section{Applications}
The proposed and existing multipliers are evaluated using two neural network applications: handwritten digit recognition on the MNIST dataset and image denoising using the FFdNet architecture. These benchmarks help assess the impact of multiplier design on inference accuracy and efficiency.

\subsection{MNIST Handwritten Digit Recognition}
\begin{figure}
    \centering
    \includegraphics[width=0.70\linewidth]{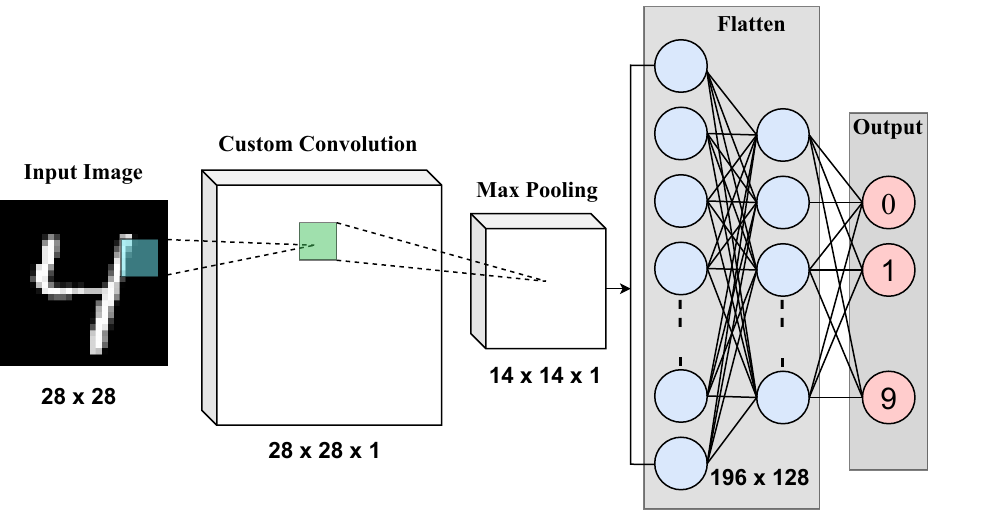}
    \caption{Architecture of Convolution Neural Network using Keras Model}
    \label{fig:cnn}
\end{figure}

The proposed multipliers was evaluated using a Keras-based \cite{chollet2015keras} convolutional neural network model (CNN), as depicted in Fig.~\ref{fig:cnn} and LeNet-5 model \cite{lecun1998gradient} developed for handwritten digit classification (0–9) using the MNIST dataset \cite{lecun1998mnist}. In this evaluation, the exact multiplier in the convolutional layers were substituted with the proposed approximate multiplier. The dataset consists of 5,000 grayscale training images and 500 testing images of handwritten digits, each with a resolution of 28×28 pixels. The CNN architecture was trained over 50 epochs, and the classification accuracies corresponding to various multiplier designs are reported in Table~\ref{tab:output_table}. Although a marginal decline in classification accuracy is observed when compared to exact multiplier, the proposed design demonstrates substantial advantages in terms of reduced power consumption and lower area overhead.
\vspace{-8mm}

\begin{table}[!h]
    \centering
    \caption{Performing Number-Recognition using different convolution Models}
    \vspace{2mm}
    \label{tab:output_table}
    \renewcommand{\arraystretch}{1.1} 
    \setlength{\tabcolsep}{5pt} 
    {
    \begin{tabular}{|c|c|c|c|}
        \hline
        \textbf{Models} & \textbf{Dataset} & \textbf{Design} & \textbf{Accuracy(\%)} \\
        \hline
        \hline
        \multirow{6}{*}{Keras\cite{chollet2015keras}}
            &       & Exact                    & 95.24 \\
            &       & Design\cite{Zhang2023}   & 90.58 \\
            &       & Design\cite{Anil2023}    & 92.14 \\
            & MNIST & Design\cite{Anupam2025}  & 92.46 \\
            &       & Design\cite{hemanth2024} & 93.19 \\
            &       & Proposed                 & 93.54 \\
        \hline
        \hline
        \multirow{6}{*}{LeNet-5 \cite{lecun1998gradient}}
            &       & Exact                    & 98.24 \\
            &       & Design\cite{Zhang2023}   & 91.66 \\
            &       & Design\cite{Anil2023}    & 93.72 \\
            & MNIST & Design\cite{Anupam2025}  & 93.88 \\
            &       & Design\cite{hemanth2024} & 95.12 \\
            &       & Proposed                 & 96.45 \\
        \hline
    \end{tabular}}
\end{table}
\vspace{-10mm}

\subsection{Image Denoising using FFDNET Architecture }

\begin{figure}
    \centering
    \includegraphics[width=1\linewidth]{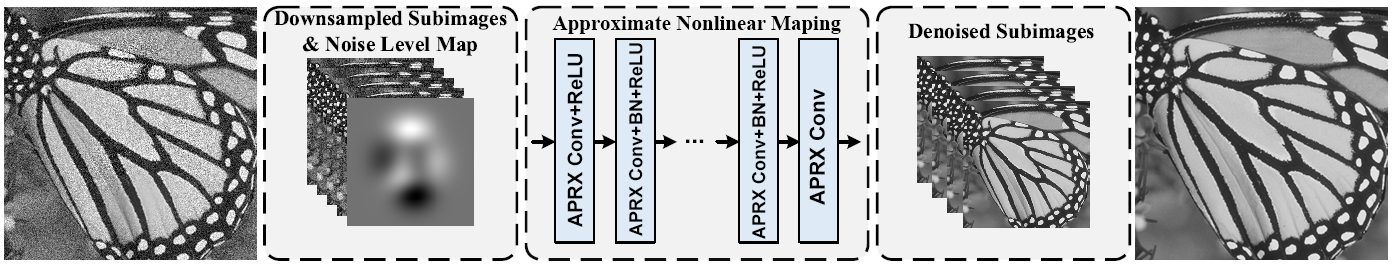}
    \caption{FFDNet Architecture with Custom Convolution Layers for Image Denoising\cite{Zhang_Kai}}
    \label{fig:FFDNet}
\end{figure}

To further evaluate the effectiveness of the proposed approximate multiplier in real-world applications, it was integrated into the convolutional layers of the FFDNet architecture, a well-established CNN model for image denoising \cite{Zhang_Kai} depicted in Fig.~\ref{fig:FFDNet}. In the original design, FFDNet employs a reversible downsampling operator followed by a sequence of convolutional layers with combinations of convolution, batch normalization, and ReLU activations to restore clean images from noisy inputs.

\begin{figure*}[t]

 \begin{tabular}{c c c c c c } 
\includegraphics[scale = 0.19]{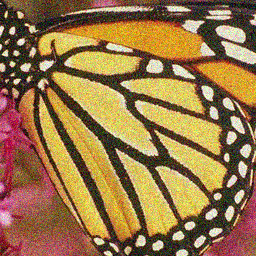} & 
\includegraphics[scale = 0.19]{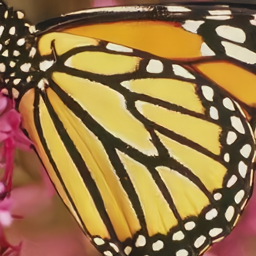}& 
\includegraphics[scale = 0.19]{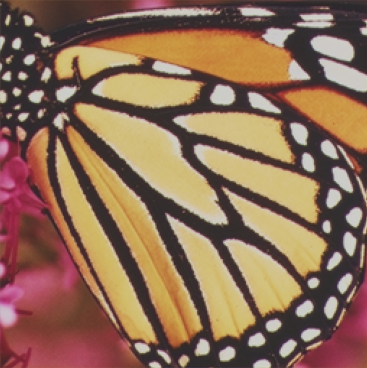}&
\includegraphics[scale = 0.19]{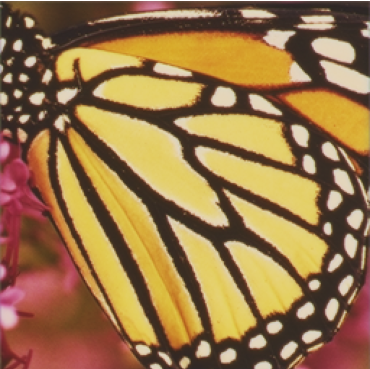} & 
\includegraphics[scale = 0.19]{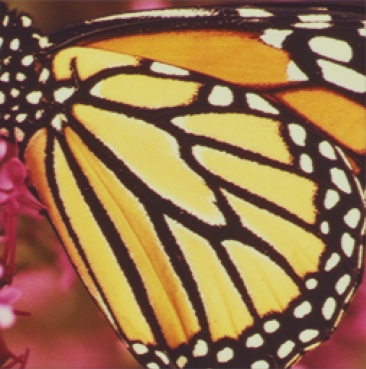}&
\includegraphics[scale = 0.19]{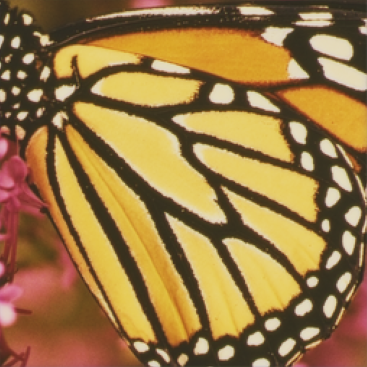}\\ 
\scriptsize{Noisy $\sigma$ = 25}  
&\scriptsize{Exact}
& \scriptsize{Design\cite{Zhang2023}} 
& \scriptsize{Design\cite{Anupam2025}}
& \scriptsize{Design\cite{hemanth2024}}  
& \scriptsize{Proposed}\\
 & \scriptsize{(35.41 dB, 0.989)}
 & \scriptsize{(30.48 dB, 0.860)}  
 & \scriptsize{(33.86 dB, 0.9402)}  
 & \scriptsize{(34.73 dB, 0.9557)}
 & \scriptsize{(34.95 dB, 0.9713)}  \\

\includegraphics[scale = 0.19]{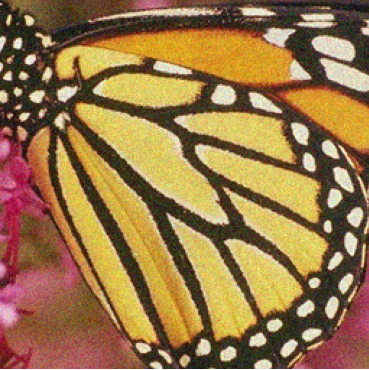}& 
\includegraphics[scale = 0.19]{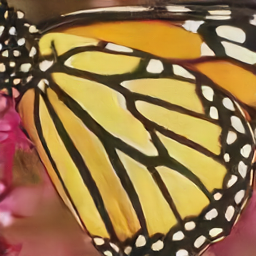}&
\includegraphics[scale = 0.19]{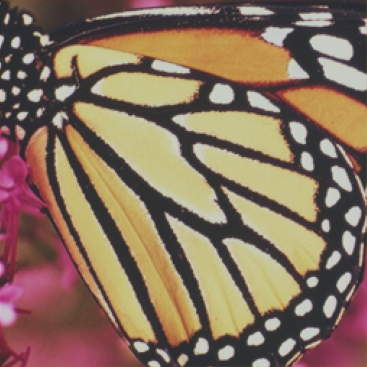} & 
\includegraphics[scale = 0.19]{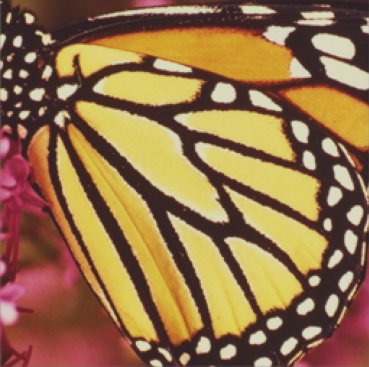}&
\includegraphics[scale = 0.19]{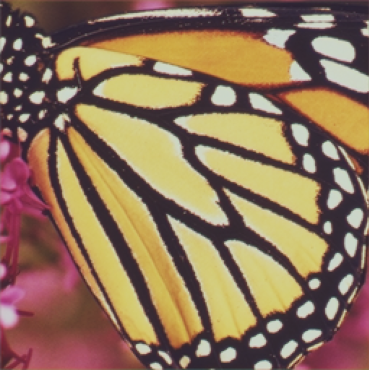} & 
\includegraphics[scale = 0.19]{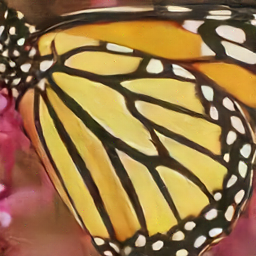}  \\
 \scriptsize{Noisy $\sigma$ = 50}  
 & \scriptsize{Denoised (Exact)}
& \scriptsize{Design\cite{Zhang2023}} 
& \scriptsize{Design\cite{Anupam2025}}
& \scriptsize{Design\cite{hemanth2024}}  
& \scriptsize{Proposed}\\
 
 & \scriptsize{(33.86 dB, 0.958)}  
 & \scriptsize{(28.80 dB, 0.791)}
 & \scriptsize{(29.25 dB, 0.852)}  
 & \scriptsize{(30.05 dB, 0.895)}  
 & \scriptsize{(32.84 dB, 0.9183)}  \\
\end{tabular}
\caption{ Denoised Output Images Using the Proposed Approximate Multiplier Integrated into the FFDNet Model, Along with Corresponding PSNR and SSIM Values for Noise Levels $\sigma$ = 25 and $\sigma$ = 50.}
\label{fig:ffdnet_results}
\end{figure*}

\begin{figure}
    \centering
    \includegraphics[width=0.4\linewidth]{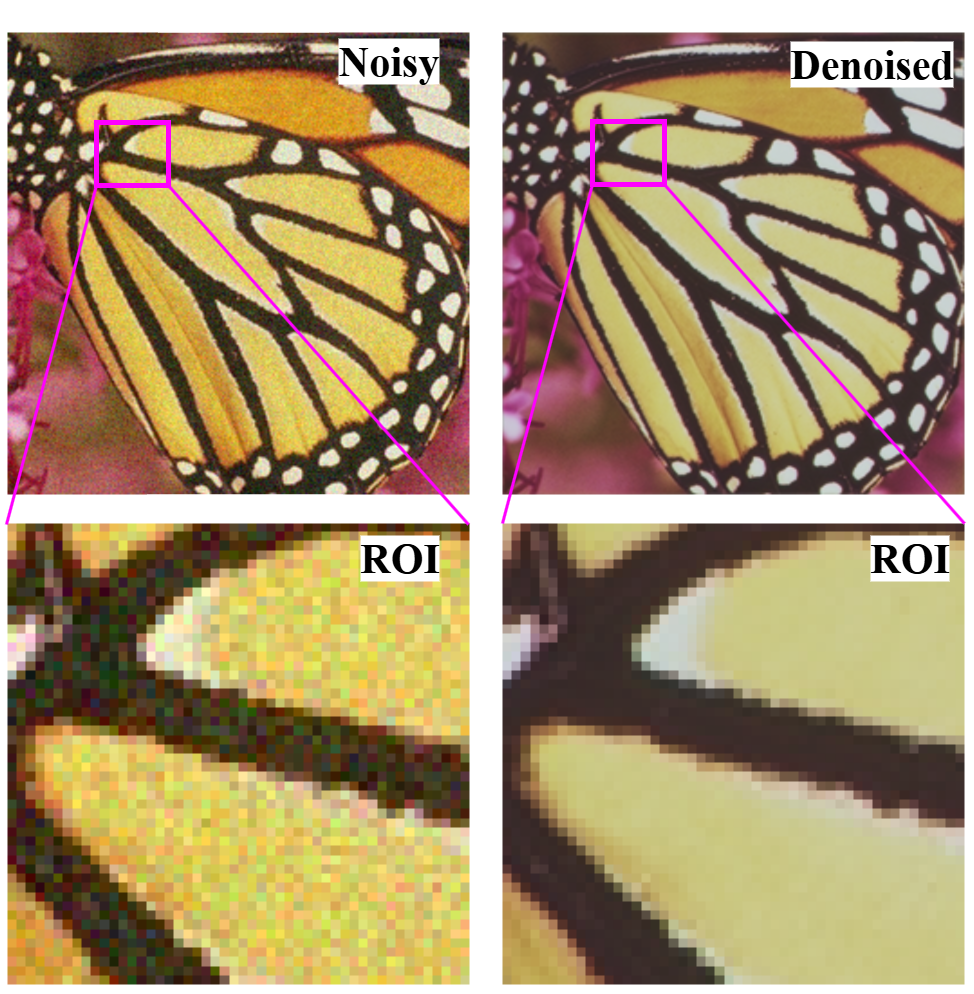}
    \caption{Comparison of Noisy and Denoised Images with Region of Interest (ROI)}
    \label{fig:Noisy and Denoised Images}
\end{figure}

In this study, the exact multiplier in the convolutional layers was substituted with the proposed approximate multiplier, while preserving the rest of the network architecture. As illustrated in Fig.~\ref{fig:Noisy and Denoised Images} and Fig.~\ref{fig:ffdnet_results}, the proposed multiplier achieves effective noise reduction with minimal perceptual degradation, retaining competitive denoising performance and high PSNR values compared to existing designs. These results highlight both the visual quality and hardware efficiency of the proposed architecture, demonstrating significant reductions in power consumption and area overhead.

\section{Conclusion}
\label{sec:conclusion}
This paper proposed a high-efficiency approximate 
8×8 unsigned multiplier incorporating a high accuracy 4:2 compressor, 
optimized for deep neural network applications. 
The architecture achieves notable accuracy and energy efficiency improvements upto 30.24\% compared to existing design, with an MRED of 0.109\% only. Experimental results demonstrate the superior performance of the proposed design in image denoising using FFDNet, achieving higher PSNR values, as well as high classification accuracy in digit recognition tasks, thereby validating its suitability for low-power AI hardware implementations.

\bibliographystyle{IEEEtran}
\bibliography{main.bib}

\end{document}